\documentclass[pra,showpacs,floatfix]{revtex4}
\usepackage{amsmath}
\usepackage{epsfig}
\usepackage{amssymb}
\usepackage{dcolumn}
\usepackage{graphicx}
\usepackage{dcolumn}
\usepackage{color}

\begin{document}

\title{Symmetric and asymmetric localized modes in linear lattices with an
embedded pair of $\chi ^{(2)}$-nonlinear sites}
\author{Valeriy A. Brazhnyi}
\email{brazhnyy@gmail.com}
\affiliation{Centro de F\'{\i}sica do Porto, Faculdade de Ci\^encias, Universidade do
Porto, R. Campo Alegre 687, Porto 4169-007, Portugal}
\author{Boris A. Malomed}
\email{malomed@post.tau.ac.il}
\affiliation{Department of Physical Electronics, School of Electrical Engineering,
Faculty of Engineering, Tel Aviv University, Tel Aviv 69978, Israel}

\begin{abstract}
We construct families of symmetric, antisymmetric, and asymmetric solitary
modes in one-dimensional bichromatic lattices with the
second-harmonic-generating ($\chi ^{(2)}$) nonlinearity concentrated at a
pair of sites placed at distance $l$. The lattice can be built as an array
of optical waveguides. Solutions are obtained in an implicit analytical
form, which is made explicit in the case of adjacent nonlinear sites, $l=1$.
The stability is analyzed through the computation of eigenvalues for small
perturbations, and verified by direct simulations. In the cascading limit,
which corresponds to large mismatch $q$, the system becomes tantamount to
the recently studied single-component lattice with two embedded sites
carrying the cubic nonlinearity. The modes undergo qualitative changes with
the variation of $q$. In particular, at $l\geq 2$, the symmetry-breaking
bifurcation (SBB), which creates asymmetric states from symmetric ones, is
supercritical and subcritical for small and large values of $q$,
respectively, while the bifurcation is always supercritical at $l=1$. In the
experiment, the corresponding change of the phase transition between the
second and first kinds may be implemented by varying the mismatch, via the
wavelength of the input beam. The existence threshold (minimum total power)
for the symmetric modes vanishes exactly at $q=0$, which suggests a
possibility to create the solitary mode using low-power beams. The stability
of solution families also changes with $q$.
\end{abstract}

\maketitle

\section{Introduction and the model}

The structure of bound states in linear systems follows the symmetry of the
underlying potential, a commonly known example being wave functions of
eigenstates in symmetric double-well potentials \cite{LL}. The addition of
the self-attractive nonlinearity leads to a qualitative change of the
situation, causing the transition from the symmetric ground state to an
asymmetric one, if the strength of the nonlinearity exceeds a critical value
\cite{misc}. This transition was studied in detail for Bose-Einstein
condensates (BECs) loaded into double-well potentials \cite{misc2}.
Experimentally, the transition was realized in BEC \cite{Markus} and in
nonlinear optics, where a double-well structure was induced in a
photorefractive material \cite{photo}.

The limit case of the double-well setting with a tall potential barrier
between the wells corresponds to the \textit{dual-core} system, such as
optical fibers \cite{Snyder} and Bragg gratings \cite{Mak} with the
twin-core structure, and pairs of linearly coupled planar waveguides with
the $\chi ^{(2)}$ (second-harmonic-generating) intrinsic nonlinearity \cite%
{Mak-chi2}. The \textit{symmetry-breaking bifurcation} (SBB), which
destabilizes the symmetric ground state in nonlinear dual-core systems and
gives rise to an asymmetric one, was first discovered in the discrete
self-trapping model \cite{Chris}. In nonlinear optics, the SBB was predicted
for continuous-wave (spatially uniform) states \cite{Snyder}, and for
solitons \cite{dual-core,Mak}, in the models of dual-core fibers and Bragg
gratings. The SBB\ was also analyzed for matter-wave solitons in the BEC
loaded into double-channel potential traps \cite{Arik}-\cite{Luca}.

The self-focusing cubic nonlinearity (the Kerr term in optics) gives rise to
the symmetry-breaking phase transition of solitons of the first kind (alias
the subcritical SBB) in the twin-core system. In that case, the branches of
asymmetric modes emerge as unstable ones, going backward and then
stabilizing themselves at turning points, from which they continue the
evolution in the forward direction \cite{bif}. The character of the phase
transition alters to the second kind (i.e., the SBB type changes from sub-
to supercritical) under the combined action of the self-focusing
nonlinearity and a periodic potential acting along the free axis (transverse
to the double-well potential) \cite{Arik,Warsaw2}. The supercritical SBB
gives rise to stable asymmetric branches which immediately go in the forward
direction \cite{bif}. The SBBs in the twin-core Bragg grating and $\chi
^{(2)}$ waveguides are of the forward type too \cite{Mak,Mak-chi2}.

A noteworthy counterpart of the linear double-well potential is a \textit{%
pseudopotential} \cite{Harrison} induced by a spatial modulation of the
nonlinearity coefficient in the form of a symmetric pair of sharp peaks.
This configuration may be implemented in optics and BEC alike \cite%
{Barcelona,Pu}. The ultimate form of the double-peak pseudopotential
features the nonlinearity concentrated at two points, in the form of a
symmetric pair of delta-functions (which may be approximated by narrow
Gaussians) \cite{Thawatchai,Nir,2012}. 
The SBB of solitons in the dual-core pseudopotentials belongs to the
subcritical type \cite{Thawatchai,Nir}. The SBB\ of localized modes was also
recently investigated in the model of a linear waveguide with two narrow $%
\chi ^{(2)}$ stripes embedded into it \cite{Asia} (the solution for the mode
pinned to a single stripe was found in Ref. [23]).

For discrete solitons in dual-core lattices, the SBB was analyzed too,
assuming the uniform transverse coupling between parallel chains \cite%
{Herring}, or the coupling applied at a single pair of sites \cite{Ljupco}.
The bifurcation is subcritical in the former case, and supercritical in the
latter one. The simplest realizations of the SBB in discrete media are
provided by linear lattices with a symmetric pair of embedded nonlinear
sites (this setting was introduced in Refs. \cite{Tsironis,BB11}), or a
symmetric pair of nonlinear elements side-coupled to the linear chain \cite%
{Almas}. In particular, the SBB for localized modes in the linear lattice
with a pair of sites carrying the cubic nonlinearity was recently studied in
Ref. \cite{BB11}, where it was concluded that the bifurcation is, chiefly,
of the subcritical type (except for the case when the nonlinear sites are
separated by a single lattice spacing, see below).

The objective of the present work is to investigate localized symmetric,
asymmetric, and antisymmetric solitary modes in the bichromatic
one-dimensional linear lattice with a pair of inserted $\chi ^{(2)}$%
-nonlinear sites. The lattice is described by the following equations:%
\begin{eqnarray}
i\frac{du_{m}}{dz}+\frac{1}{2}\left( u_{m+1}+u_{m-1}-2u_{m}\right) +\left(
\delta _{m,0}+\delta _{m-l,0}\right) u_{m}^{\ast }v_{m} &=&0,  \label{u1} \\
2i\frac{dv_{m}}{dz}+\frac{1}{2}\left( v_{m+1}+v_{m-1}-2v_{m}\right) -qv_{m}+%
\frac{1}{2}\left( \delta _{m,0}+\delta _{m-l,0}\right) u_{m}^{2} &=&0,
\label{v1}
\end{eqnarray}%
where $u_{m}$ and $v_{m}$ are complex amplitudes of the
fundamental-frequency (FF) and second-harmonic (SH) fields at the $m$-th
site, the constant of the linear coupling between adjacent sites is scaled
to be $1$, as well as the strength of the $\chi ^{(2)}$ nonlinearity
inserted at sites $m=0$ and $m=l$, which are represented by the Kronecker's
symbols $\delta _{m,0}$ and $\delta _{m-l,0}$ (i.e., integer $l$ is the
distance between the nonlinear sites), and $q$ is the real mismatch
parameter. The system can be implemented as an array of parallel optical
waveguides, with $z$ being the propagation distance along the waveguides, as
previously done for a great variety of models supported by such
quasi-discrete settings in nonlinear optics \cite{big-review}. Arrayed $\chi
^{(2)}$ structures may be also built by means of the quasi-phase-matched
technique \cite{QPM}. In these contexts, discrete $\chi ^{(2)}$ systems were
introduced in Ref. \cite{discrete-chi2}.

To realize the present model, based on Eqs. (\ref{u1}), (\ref{v1}), two
selected cores in the waveguiding array can be made quadratically nonlinear
by fabricating them of an appropriate material, or applying the $\chi ^{(2)}$%
-inducing poling to this pair \cite{Asia}. In fact, the model with two
Kerr-nonlinear sites, studied in Ref. \cite{BB11}, corresponds to the
\textit{cascading limit} \cite{Torner} of Eqs. (\ref{u1}), (\ref{v1}) for $%
q\rightarrow +\infty $. In this work, we demonstrate that the system with
negative, zero, and positive values of the mismatch ($q$) opens new
possibilities for the creation and manipulations of discrete solitary modes,
in comparison with the cascading limit. In particular, we demonstrate that
the character of the SBB can be switched from super- to subcritical by
varying the mismatch, which also alters the stability of the modes and their
existence thresholds.

The paper is organized as follows. Analytical solutions for symmetric,
asymmetric, and antisymmetric localized modes generated by Eqs. (\ref{u1}), (%
\ref{v1}) in the infinite lattice are produced in Section 2. In the general
case, the solution is implicit, while explicit solutions are obtained for
the smallest separation between the nonlinear sites, $l=1$. Numerical
results, obtained for finite lattices, are reported in Section 3. The
stability of the discrete modes is considered in that section too. The paper
is concluded by Section 4.

\section{Analytical considerations}

\subsection{The general case}

Analytical solutions to Eqs. (\ref{u1}), (\ref{v1}) for stationary modes
with propagation constant $k$ are sought for in the usual form,%
\begin{equation}
\left\{ u_{m,}v_{m}\right\} =\left\{ e^{ikz}U_{m},e^{2ikz}V_{m}\right\} ,
\end{equation}%
which reduces Eqs. (\ref{u1}) and (\ref{v1}) into the stationary equations
for real discrete fields $U_{m},V_{m}$:%
\begin{eqnarray}
-kU_{m}+\frac{1}{2}\left( U_{m+1}+U_{m-1}-2U_{m}\right) +\left( \delta
_{m,0}+\delta _{m-l,0}\right) U_{m}V_{m} &=&0,  \label{U} \\
-\left( 4k+q\right) V_{m}+\frac{1}{2}\left( V_{m+1}+V_{m-1}-2V_{m}\right) +%
\frac{1}{2}\left( \delta _{m,0}+\delta _{m-l,0}\right) U_{m}^{2} &=&0.
\label{V}
\end{eqnarray}%
At $m\leq 0$, an exact solution to Eqs. (\ref{U}) and (\ref{V}) is obvious:%
\begin{equation}
U_{m}=A_{1}\exp \left( -\kappa _{1}|m|\right) ,~V_{m}=A_{2}\exp \left(
-\kappa _{2}|m|\right) ,  \label{A12}
\end{equation}%
where $A_{1}$ and $A_{2}$ are arbitrary amplitudes, and $\kappa _{1,2}>0$
are determined by the following relations:%
\begin{equation}
k=2\sinh ^{2}\left( \kappa _{1}/2\right) ,\quad 4k+q=2\sinh ^{2}\left(
\kappa _{2}/2\right) .  \label{kappa}
\end{equation}%
Due to condition $\kappa _{1,2}>0$, the limit value of $k$ which corresponds
to $\kappa _{1}=0$ or $\kappa _{2}=0$ can be found from Eq. (\ref{kappa}):
\begin{equation}
k_{\lim }=\left\{
\begin{array}{cc}
0, & \mathrm{for}\ q>0, \\
-q/4, & \mathrm{for}\ q<0,%
\end{array}%
\right.   \label{klim}
\end{equation}%
the localized modes existing at $k\geq k_{\lim }$. Similarly, at $m\geq l$,
the exact solution to Eqs. (\ref{U}) and (\ref{V}) is%
\begin{equation}
U_{m}=C_{1}\exp \left( -\kappa _{1}(m-l)\right) ,~V_{m}=C_{2}\exp \left(
-\kappa _{2}(m-l)\right) ,  \label{C12}
\end{equation}%
and in the inner region, $0\leq m\leq l$, the solution is constructed as%
\begin{eqnarray}
U_{m} &=&B_{11}\exp \left( -\kappa _{1}m\right) +B_{12}\exp \left( -\kappa
_{1}\left( l-m\right) \right) ,~  \notag \\
V_{m} &=&B_{21}\exp \left( -\kappa _{2}m\right) +B_{22}\exp \left( -\kappa
_{2}\left( l-m\right) \right) .  \label{inner}
\end{eqnarray}

The conditions of the continuity of the discrete fields at points $m=0$ and $%
m=l$ lead to linear relations between the amplitudes of the solutions in the
outer and inner regions:%
\begin{eqnarray}
B_{11}+B_{12}\exp \left( -\kappa _{1}l\right)  &=&A_{1},~B_{11}\exp \left(
-\kappa _{1}l\right) +B_{12}=C_{1},  \notag \\
B_{21}+B_{22}\exp \left( -\kappa _{2}l\right)  &=&A_{2},~B_{21}\exp \left(
-\kappa _{2}l\right) +B_{22}=C_{2}.  \label{lin}
\end{eqnarray}%
Finally, the nonlinear equations at sites $m=0$ and $m=l$ take the following
form:%
\begin{gather}
-\left[ k+1-\frac{1}{2}\exp \left( -\kappa _{1}\right) \right] A_{1}  \notag
\\
+\frac{1}{2}\left[ B_{11}\exp \left( -\kappa _{1}\right) +B_{12}\exp \left(
-\left( l-1\right) \kappa _{1}\right) \right] +A_{1}A_{2}=0,  \notag \\
-\left[ 4k+q+1-\frac{1}{2}\exp \left( -\kappa _{2}\right) \right] A_{2}
\notag \\
+\frac{1}{2}\left[ B_{21}\exp \left( -\kappa _{2}\right) +B_{22}\exp \left(
-\left( l-1\right) \kappa _{2}\right) \right] +\frac{1}{2}A_{1}^{2}=0,
\label{m=0}
\end{gather}%
\begin{gather}
-\left[ k+1-\frac{1}{2}\exp \left( -\kappa _{1}\right) \right] C_{1}  \notag
\\
+\frac{1}{2}\left[ B_{11}\exp \left( -\kappa _{1}\left( l-1\right) \right)
+B_{12}\exp \left( -\kappa _{1}\right) \right] +C_{1}C_{2}=0,  \notag \\
-\left[ 4k+q+1-\frac{1}{2}\exp \left( -\kappa _{2}\right) \right] C_{2}
\notag \\
+\frac{1}{2}\left[ B_{21}\exp \left( -\kappa _{2}\left( l-1\right) \right)
+B_{22}\exp \left( -\kappa _{2}\right) \right] +\frac{1}{2}C_{1}^{2}=0.
\label{1}
\end{gather}%
Amplitudes $B_{11,12}$ and $B_{21,22}$ can be eliminated in favor of $A_{1,2}
$ and $C_{1,2}$, making use of linear equations (\ref{lin}), which leads to
the system of four equations, (\ref{m=0}) and (\ref{1}), for four remaining
unknown amplitudes, $A_{1,2}$ and $C_{1,2}$. Families of solutions for
localized modes are characterized below by dependences of their powers
(norms),
\begin{equation}
N_{\left\{ U,V\right\} }=\sum_{m=-\infty }^{m=+\infty }\left\{
U_{m}^{2},V_{m}^{2}\right\} ,  \label{N}
\end{equation}%
on the propagation constant, $k$, a dynamical invariant of Eqs. (\ref{u1}), (%
\ref{v1}) being the total power,
\begin{equation}
N=N_{U}+4N_{V}.  \label{total}
\end{equation}

The solutions for the symmetric and antisymmetric modes are defined,
respectively, by constraints $A_{n}=C_{n}$, $B_{n1}=B_{n2}$, or $%
A_{n}=(-1)^{n}C_{n}$, $B_{n1}=(-1)^{n}B_{n2},$ with $n=1,2$. In the latter
case, only the FF field is antisymmetric, while its SH counterpart remains
symmetric; note also that the antisymmetric modes with even and odd $l$ may
be categorized, respectively, as ones of the on-site and inter-site types
\cite{BB11}. Explicit solutions for the symmetric and antisymmetric modes
can be obtained, in a simple approximate form, for $0\leq k,4k+q\ll 1$, when
Eq. (\ref{kappa}) yields
\begin{equation}
\kappa _{1}\approx \sqrt{2k},\kappa _{2}\approx \sqrt{2\left( 4k+q\right) },
\label{kappakappa}
\end{equation}%
hence the modes are broad in this limit, according to Eq. (\ref{A12}).
Further, the amplitudes and powers (\ref{N}) of the symmetric and
antisymmetric modes, found in the same limit, are%
\begin{eqnarray}
\mathrm{symm}\mathrm{:~} &&A_{1}^{2}\approx \kappa _{2}A_{2}\approx
(1/2)\kappa _{1}\kappa _{2},~N_{U}\approx \kappa _{2}/2,~N_{V}\approx \kappa
_{1}^{2}/\left( 4\kappa _{2}\right) ,  \notag \\
\mathrm{antisymm}\mathrm{:~} &&A_{1}^{2}\approx \kappa _{2}A_{2}\approx
\kappa _{2}/l,~N_{U}\approx \kappa _{2}/\left( l\kappa _{1}\right)
,~N_{V}\approx \left( l^{2}\kappa _{2}\right) ^{-1}.  \label{sas}
\end{eqnarray}%
A corollary of this result is that, except for the case of $q=0$, when $%
\kappa _{2}\approx 2\kappa _{1},$ the total power of the modes never
vanishes at $\kappa _{1,2}\rightarrow 0$, hence there is a finite power
threshold (a minimum value of the total power) necessary for the existence
of the localized states at $q\neq 0$. In particular, for the symmetric modes
with $0\leq q\ll 1$, the threshold is $N_{\min }=N_{U}(k=\kappa
_{1}=0)\approx \sqrt{q/2}$, which, indeed, vanishes solely at $q=0$.

On the other hand, in lattices of a finite size, the threshold power of the
symmetric modes vanishes in the limit of $\kappa _{1}\approx \sqrt{2k}%
\rightarrow 0$ [i.e., when $\kappa _{2}$ remains finite, provided that $q$
is positive, see Eq. (\ref{kappakappa})]. Indeed, in this limit case the top
line in Eq. (\ref{sas}) implies the vanishing of both $A_{1}$ and $A_{2}$,
while the nonvanishing of the threshold power in the respective infinite
lattice is accounted for by the simultaneous divergence of the spatial width
of the FF component, $\kappa _{1}^{-1}\rightarrow \infty $. Obviously, the
latter factor cannot compensate the vanishing of the amplitude in
finite-size lattices.

\subsection{The case of $l=1$: Explicit results}

The above analysis yields results in the implicit form, given by Eqs. (\ref%
{lin})-(\ref{1}). Explicit solutions can be obtained for the smallest
distance between the nonlinear sites, $l=1$. In this case, one does not need
to introduce solutions (\ref{inner}) for the inner layer, while the
remaining equations for amplitudes $A_{1,2}$ and $C_{1,2}$ take the
following form:%
\begin{eqnarray}
-\left[ k+1-\frac{1}{2}\exp \left( -\kappa _{1}\right) \right] A_{1}+\frac{1%
}{2}C_{1}+A_{1}A_{2} &=&0,  \label{11-1} \\
-\left[ 4k+q+1-\frac{1}{2}\exp \left( -\kappa _{2}\right) \right] A_{2}+%
\frac{1}{2}C_{2}+\frac{1}{2}A_{1}^{2} &=&0,  \label{12-1}
\end{eqnarray}%
\begin{eqnarray}
-\left[ k+1-\frac{1}{2}\exp \left( -\kappa _{1}\right) \right] C_{1}+\frac{1%
}{2}A_{1}+C_{1}C_{2} &=&0,  \label{21-1} \\
-\left[ 4k+q+1-\frac{1}{2}\exp \left( -\kappa _{2}\right) \right] C_{2}+%
\frac{1}{2}A_{2}+\frac{1}{2}C_{1}^{2} &=&0.  \label{22-1}
\end{eqnarray}%
The symmetric solution, with $A_{1}=C_{1}$ and $A_{2}=C_{2}$, can be easily
found from here:%
\begin{eqnarray}
A_{2} &=&k+\frac{1}{2}-\frac{1}{2}\exp \left( -\kappa _{1}\right) ,  \notag
\\
A_{1}^{2} &=&2\left[ 4k+q+\frac{1}{2}-\frac{1}{2}\exp \left( -\kappa
_{2}\right) \right] A_{2}.  \label{symm-1}
\end{eqnarray}%
The search for the SBB, i.e., solutions with infinitely small $A_{1}-C_{1}$
and $A_{2}-C_{2}$, yields an equation for the bifurcation point: $%
A_{1}^{2}=4k+q+1-\frac{1}{2}\exp \left( -\kappa _{2}\right) $. Substituting $%
A_{1}^{2}$ from Eq. (\ref{symm-1}), it takes the following form:%
\begin{equation}
\left[ 4k+q+\frac{1}{2}-\frac{1}{2}\exp \left( -\kappa _{2}\right) \right] %
\left[ 2k-\exp \left( -\kappa _{1}\right) \right] =1.  \label{bif-1}
\end{equation}

Further analysis demonstrates that, at $l=1$, the SBB is supercritical at
all values of $q$, including the cascading limit, $q\rightarrow \infty $.
The latter fact implies that the SBB is also supercritical in the model with
the cubic nonlinearity at two adjacent sites. Indeed, this can be checked to
be correct in the cubic model, while for all $l\geq 2$ the corresponding SBB
is subcritical \cite{BB11} (the character of the SBB for $l=1$, super- or
subcritical, was not considered in Ref. \cite{BB11}).

It is also possible to find antisymmetric solutions, with $A_{1}=-C_{1}$, $%
A_{2}=C_{2}$:
\begin{eqnarray}
A_{2} &=&k+\frac{3}{2}-\frac{1}{2}\exp \left( -\kappa _{1}\right) ,  \notag
\\
A_{1}^{2} &=&2\left[ 4k+q+\frac{1}{2}-\frac{1}{2}\exp \left( -\kappa
_{2}\right) \right] A_{2}.  \label{anti-1}
\end{eqnarray}%
A point where the antisymmetry could be spontaneously broken, similar to the
SBB of the symmetric modes, is determined by the condition that solutions
with infinitely small $A_{1}+C_{1}$ and $A_{2}-C_{2}$ emerge. After a simple
analysis, this condition leads to the following equation:%
\begin{equation}
\left[ 4k+q+\frac{1}{2}-\frac{1}{2}\exp \left( -\kappa _{2}\right) \right] %
\left[ 2k+2-\exp \left( -\kappa _{1}\right) \right] =1,  \label{antibif}
\end{equation}%
cf. Eq. (\ref{bif-1}). Replacing here $4k+q$ and $k$ in the square-bracket
combinations by their expressions in terms of $\kappa _{2}$ and $\kappa _{1}$
given by Eqs. (\ref{kappa}), it is easy to check that Eq. (\ref{antibif})
can \emph{never} be satisfied, unlike Eq. (\ref{bif-1}) (the left-hand side
always takes values $\geq 1$), hence the antisymmetric modes \emph{do not}
undergo the bifurcation.


\section{Numerical results}

\subsection{Zero mismatch ($q=0$)}

\begin{figure}[th]
\epsfig{file=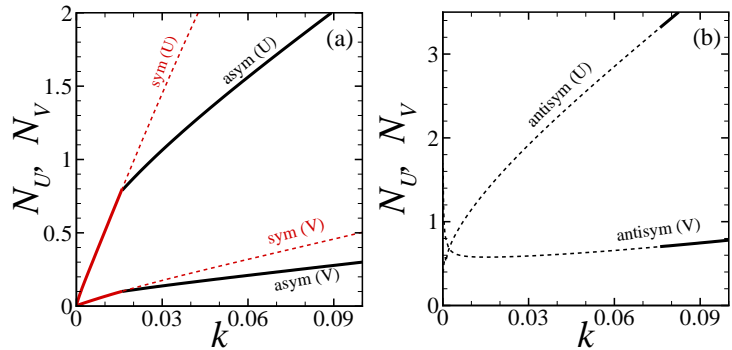,width=10cm} 
\caption{(Color online) The powers (norms) of the FF ($U$) and SH$\ $($V$)
components of one-dimensional localized modes versus the propagation
constant, as produced by the implicit analytical solution for the infinite
lattice, based on Eqs. (\protect\ref{lin})-(\protect\ref{1}), and by the
numerical solution for the finite lattice with two nonlinear sites,
separated by distance $l=5$, with zero mismatch, $q=0$. The analytical and
numerical curves completely coincide. Families of symmetric and asymmetric
solutions, and antisymmetric ones, are shown, severally, in panels (a) and
(b). Solid and dashed lines correspond to the stable and unstable branches,
respectively. }
\label{fig1}
\end{figure}

First, we present families of numerical solutions of Eqs. (\ref{U}) and (\ref%
{V}) with a finite number of sites, $\mathcal{N}=91$, obtained for $q=0$ and
$l=5$. In Fig. \ref{fig1}, the families of symmetric, asymmetric, and
antisymmetric modes are represented by the $N_{U,V}(k)$ curves for the FF
and SH components. These curves completely overlap with those predicted by
the implicit analytical solutions, in the form of Eqs. (\ref{lin})-(\ref{1}).

In Fig. \ref{fig1}(a), the family of symmetric solutions features the SBB at
a finite value of $k$, and an asymptotic behavior with $N_{U,V}\rightarrow 0$
at $k\rightarrow 0$ (i.e., the vanishing of the threshold power), in
accordance with the analytical approximation based on Eqs. (\ref{kappakappa}%
) and (\ref{sas}). This result is of obvious interest to the
experiment, suggesting the possibility to create the nonlinear
guided modes, using low-power input beams. On the other hand, Fig.
\ref{fig1}(b) demonstrates a completely different behavior of the
norms of the antisymmetric solutions at
$k\rightarrow 0$: While $N_{U}$ approaches a finite value in this limit, $%
N_{V}$ diverges at $k\rightarrow 0$, also in accordance with Eqs. (\ref{sas}%
) and (\ref{kappakappa}) (in terms of the numerical results, both the
divergence of $N_{V}$ and maintaining the finite value of $N_{U}$ are
limited by the finite size of the lattice).

The stability of the localized modes of the different types, which is also
indicated in Fig. \ref{fig1}, was established by solving the linear
eigenvalue problem for small perturbations, and verified through direct
simulations of perturbed evolution in the framework of Eqs. (\ref{u1}), (\ref%
{v1}).

\begin{figure}[th]
\epsfig{file=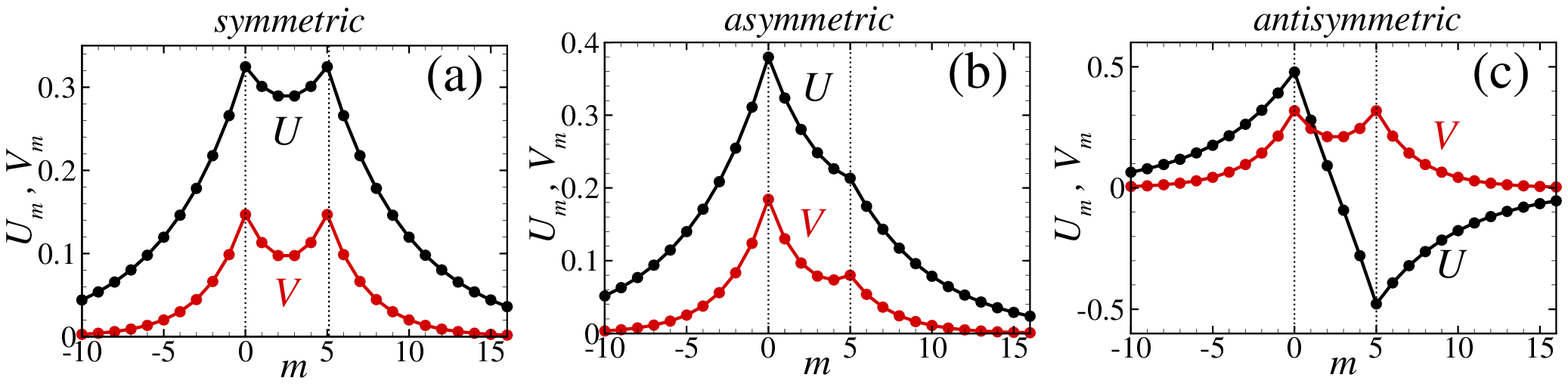,width=12cm}
\caption{(Color online) Examples of analytical (solid line) and numerical
(points) solutions for symmetric (unstable, in this case), asymmetric
(stable), and antisymmetric (unstable) modes, found for $k=0.02$, $l=5$ and $%
q=0$. Black and red colors show the $U$- and $V$-fields, respectively.}
\label{fig2}
\end{figure}

\begin{figure}[th]
\epsfig{file=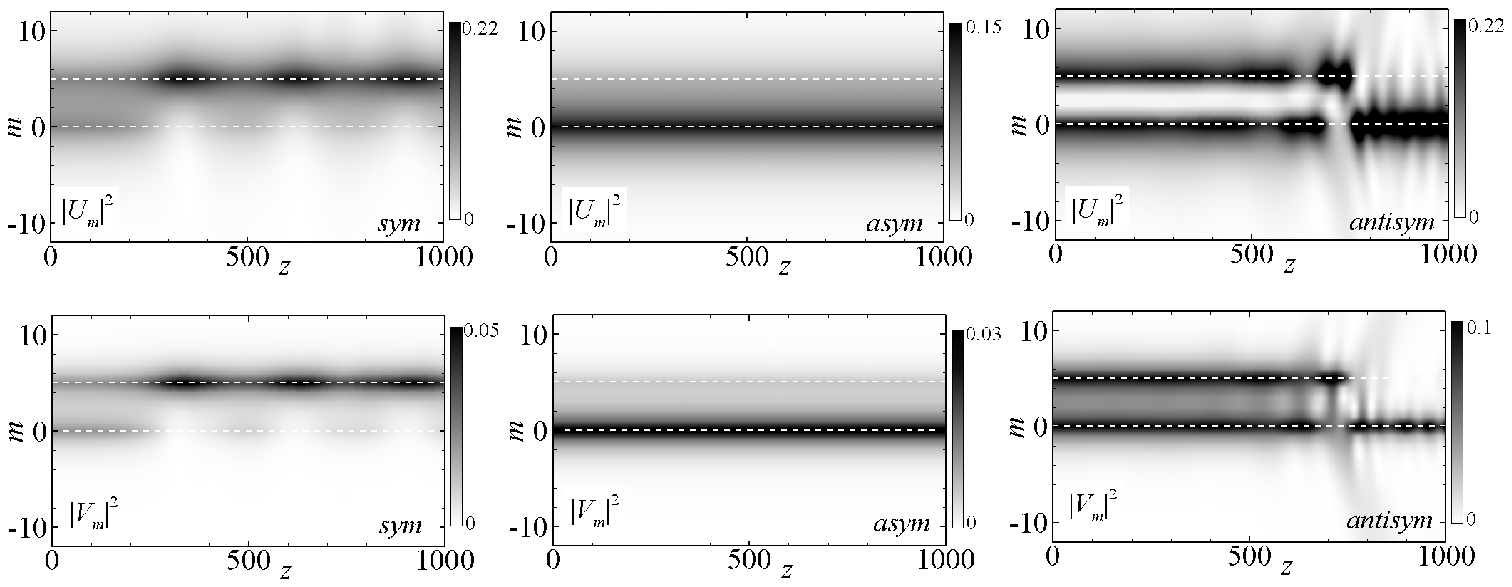,width=12cm}%
\epsfig{file=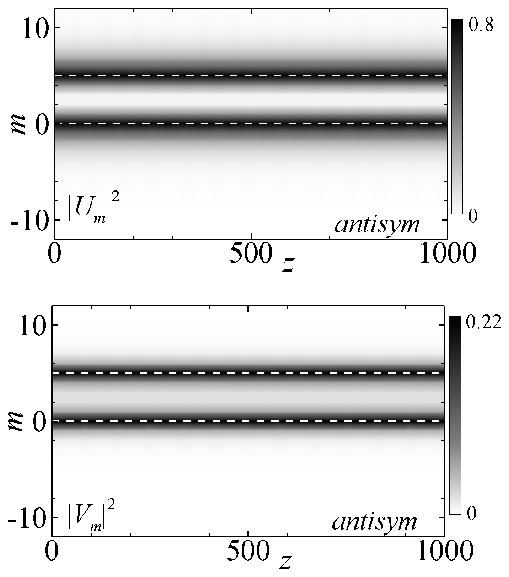,width=4cm}
\caption{(Color online) The first three columns display density plots
representing the perturbed evolution of symmetric (unstable), asymmetric
(stable) and antisymmetric (unstable) solutions from Fig. \protect\ref{fig2}%
. The fourth column displays the density plot of the stable antisymmetric
solution at $k=0.08$.}
\label{fig2_dyn}
\end{figure}

Typical profiles of modes featuring the different types of the symmetry are
shown in Fig. \ref{fig2}, and the (in)stability of these modes is
illustrated in the first three columns of Figs. \ref{fig2_dyn} by means of
direct simulations. As might be expected, the instability of the symmetric
localized modes past the SBB point ($k_{\mathrm{SBB}}\approx 0.015$)
transforms it into an excited state close to a stable asymmetric mode. The
simulations also corroborate the change of the stability of the
antisymmetric modes [at $k\approx 0.075$ in Fig. \ref{fig1}(b)], which is
predicted by the computation of eigenvalues for small perturbations. In
particular, the fourth column of Fig. \ref{fig2_dyn} demonstrates an example
of stable antisymmetric modes.

\begin{figure}[th]
\epsfig{file=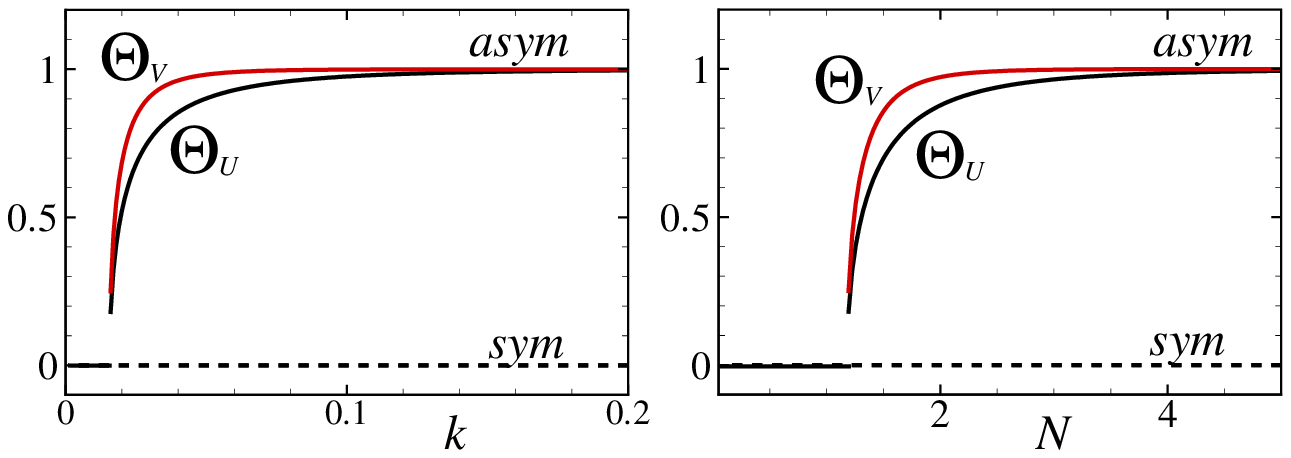,width=10cm}
\caption{ (Color online) The dependence of asymmetry measures (\protect\ref%
{theta}) on propagation constant $k$ and total power $N$ in the case of the
supercritical bifurcation, at $l=5$ and $q=0$. Black and red colors show the
$U$- and $V$-fields, respectively. }
\label{fig5}
\end{figure}

To highlight the character of the SBB, we use natural measures of the
asymmetry of the solutions taken in the form of Eqs. (\ref{A12})-(\ref{inner}%
), namely,
\begin{equation}
\Theta _{U}\equiv \frac{A_{1}^{2}-C_{1}^{2}}{A_{1}^{2}+C_{1}^{2}},~\Theta
_{V}\equiv \frac{A_{2}^{2}-C_{2}^{2}}{A_{2}^{2}+C_{2}^{2}}.  \label{theta}
\end{equation}%
For $q=0$, the asymmetries are plotted versus $k$ and the total power
(norm), defined as per Eq. (\ref{total}), in Fig. \ref{fig5}, which
demonstrates that the bifurcation is supercritical in this case.

\subsection{Negative mismatch ($q<0$)}

\begin{figure}[th]
\epsfig{file=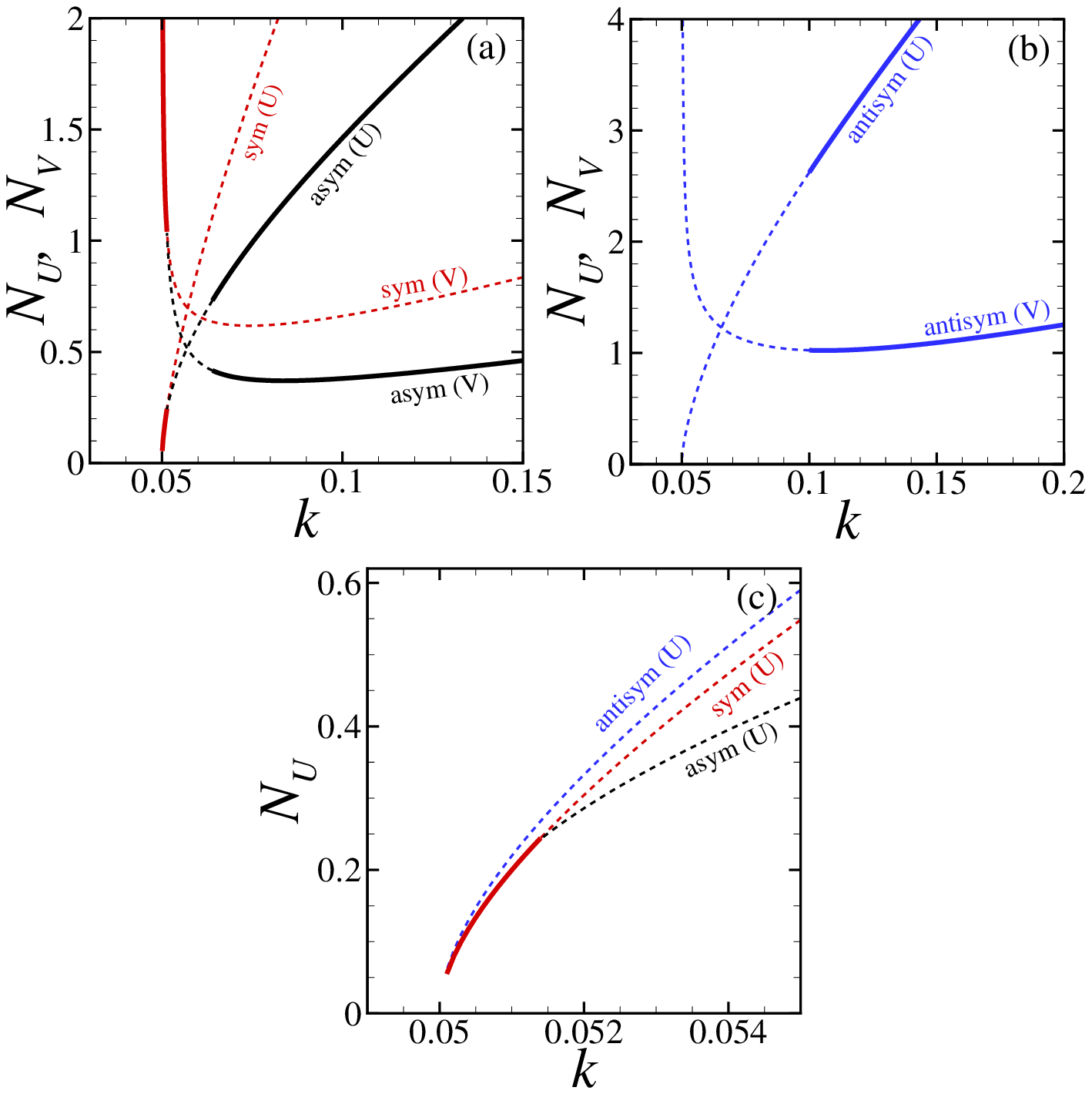,width=12cm}
\caption{(Color online) Panels (a) and (b) show the same as in Fig. \protect
\ref{fig1}, but for $q=-0.2$. Panel (c) is a blowup of the bifurcation
region for the $U$-field.}
\label{fig3}
\end{figure}

A behavior different from that at $q=0$ is observed at $q<0$. As seen in
Fig. \ref{fig3}, not only the antisymmetric solutions but also symmetric
ones feature the divergence of the norm of the $V$-field as $k$ approaches
the corresponding limit value $k_{\lim }=-q/4,$ see Eq. (\ref{klim}). This
feature can also be easily explained by the above analysis: In the limit of $%
k+q/4\rightarrow 0$, Eq. (\ref{kappa}) yields $\kappa _{2}\approx \sqrt{%
2\left( 4k+q\right) }$, hence the respective width $\sim \kappa _{2}^{-1}$
diverges, while $\kappa _{1}$ remains finite [cf. Eq. (\ref{kappakappa})].
Further, Eqs. (\ref{lin})-(\ref{1}) yield, in this limit, finite $A_{2}$ for
both symmetric and antisymmetric branches, while the respective value of $%
A_{1}$ is vanishing as per $A_{1}^{2}\approx \kappa _{2}A_{2}$ cf. Eq. (\ref%
{sas})], hence the corresponding norm $N_{U}$ vanishes too, while $%
N_{V}\approx \kappa _{2}^{-1}A_{2}^{2}$ diverges.

\begin{figure}[th]
\epsfig{file=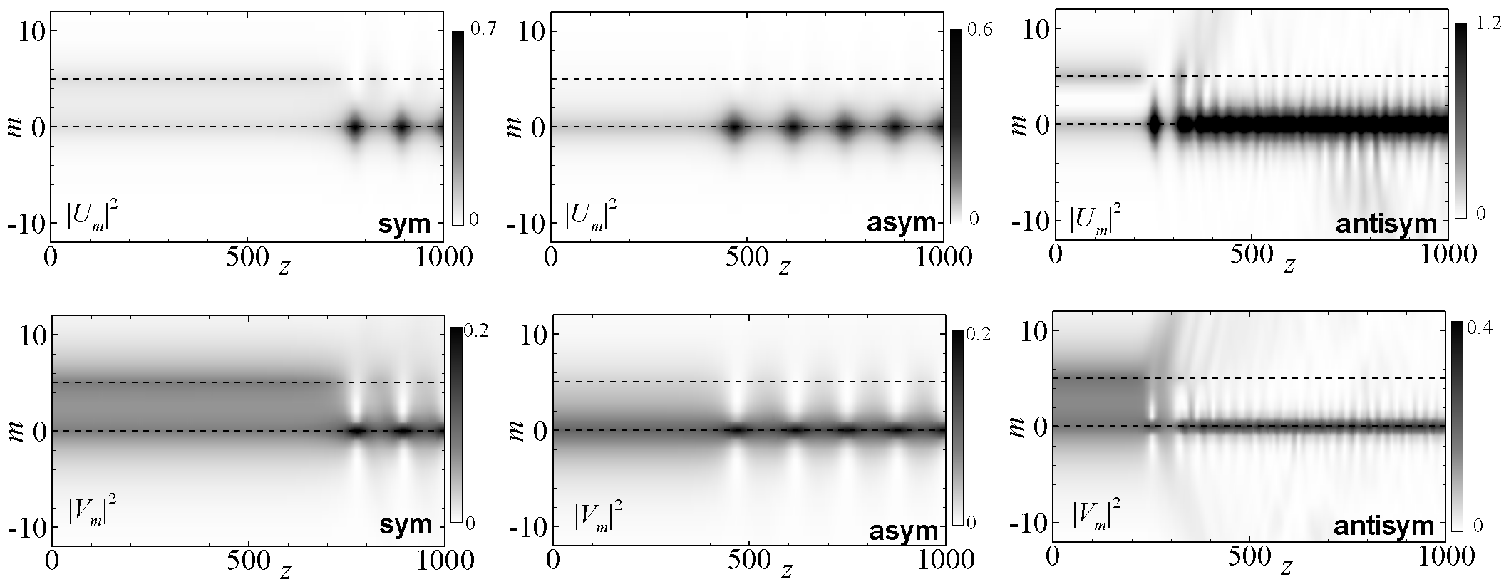,width=12cm}
\caption{(Color online) The evolution of symmetric, asymmetric and
antisymmetric unstable modes at $k=0.054$, for $l=5$ and $q=-0.2$. }
\label{fig3_dyn}
\end{figure}

The linear-stability analysis demonstrates that the symmetric branch is
stable up to the SBB point ($k_{\mathrm{SBB}}\approx 0.0514$). Past the
bifurcation point, the asymmetric branch appears in an unstable form, and
with the further increase of $k$ it becomes stable at $k\approx 0.064$.
Direct numerical simulations displayed in Fig. \ref{fig3_dyn} confirm the
predictions of the linear-stability analysis. In particular, the unstable
asymmetric and antisymmetric modes are transformed by the perturbed
evolution into breathers.

Finally, the SBB in the case of $q<0$ is of the supercritical type, similar
to that displayed in Fig. \ref{fig5} for $q=0$.

\subsection{Positive mismatch ($q>0$)}

\begin{figure}[th]
\epsfig{file=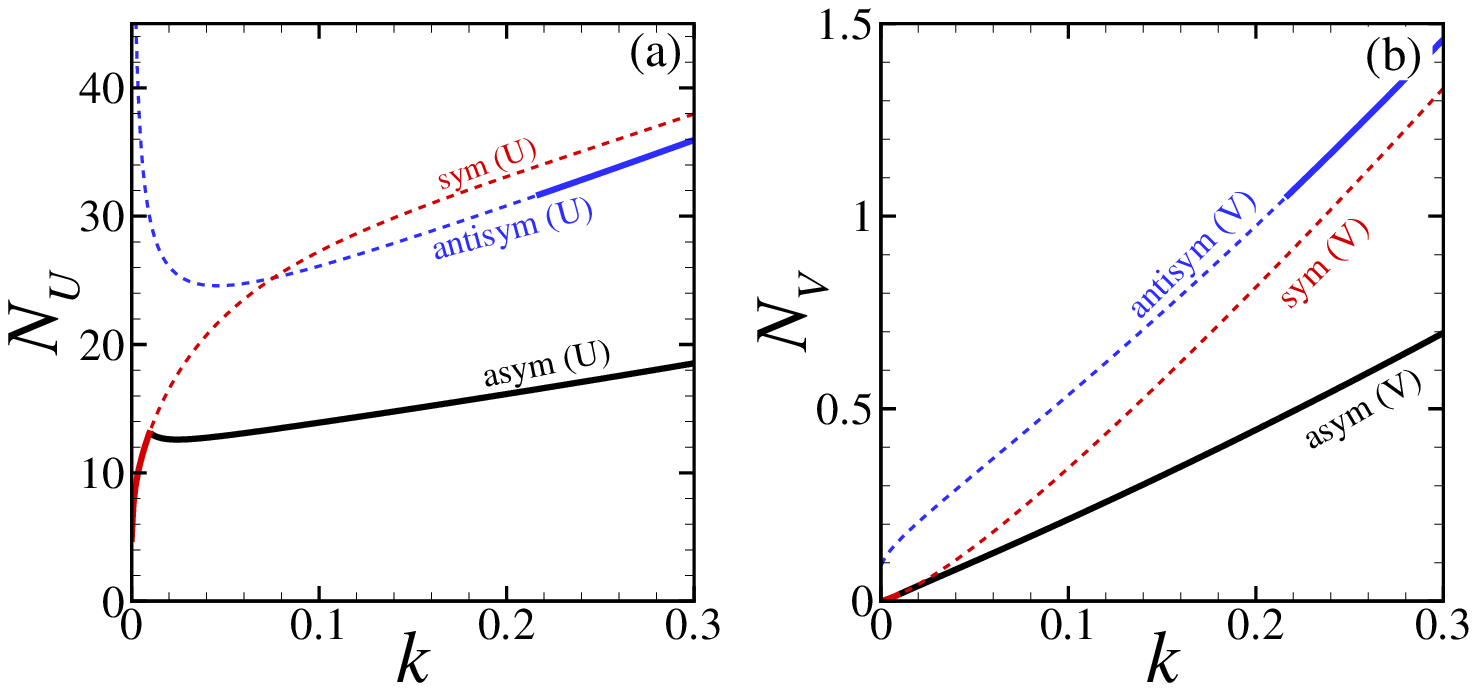,width=12cm}
\caption{(Color online) The SBB diagram of the localized modes, as produced
by the analytical solution for the infinite lattice with two nonlinear
sites, separated by distance $l=5$, and mismatch $q=5$ (the numerical
solution for the same case is indistinguishable from the analytical one).
Solid and dashed lines correspond to stable and unstable branches,
respectively.}
\label{fig4}
\end{figure}

\begin{figure}[th]
\epsfig{file=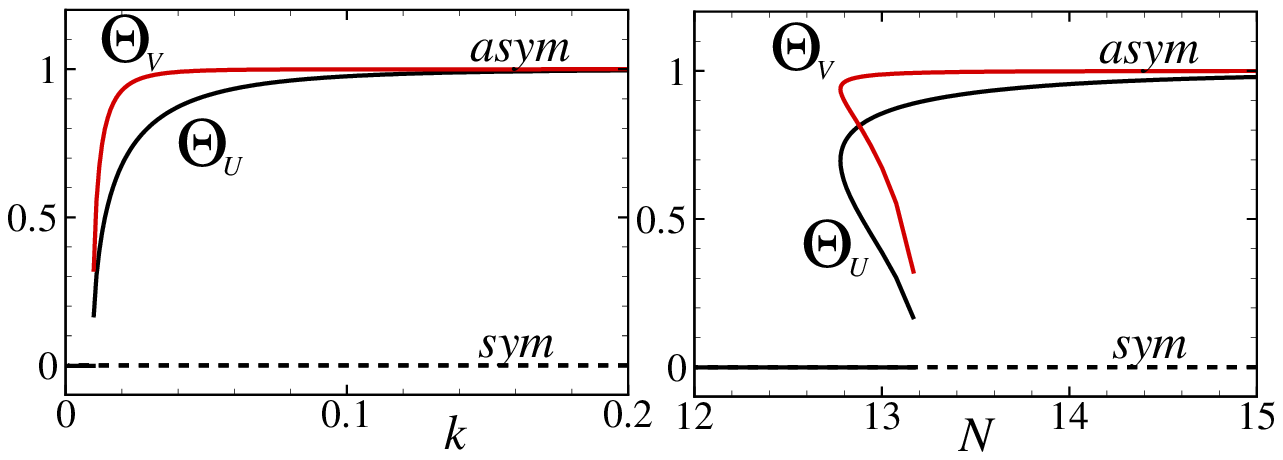,width=10cm}
\caption{(Color online) The same as in Fig. \protect\ref{fig5}, but for $q=5$%
, with the respective bifurcation being subcritical.}
\label{fig6}
\end{figure}

The increase of mismatch $q$ to large positive values leads to a transition
between the supercritical and subcritical types of the SBB. In particular,
Fig. \ref{fig4} shows that the $N_{U}(k)$ branch of the asymmetric solutions
passes a shallow minimum right after the bifurcation point, which is a
signature of a subcritical bifurcation, while the bifurcation in Fig. \ref%
{fig1}(a) was supercritical. The change of the type of the SBB at $q>q_{0}$
from super- to subcritical is further illustrated by Fig. \ref{fig6}. It is
worth noting that the bifurcation is subcritical only in terms of the FF
field, $U$.

For all $l\geq 2$ it is possible to find a boundary value $q_{0}$, such that
the SBB is of the super- and subcritical types at $q<q_{0}$ and $q>q_{0}$,
respectively. Numerical analysis of Eqs. (\ref{m=0}) and (\ref{1})
demonstrates that $q_{0}$ rapidly grows with the decrease of $l$: $%
q_{0}(l=5)\approx 0.27$, $q_{0}(l=3)\approx 1.86$, and $q_{0}(l=2)\approx 4.9
$. As said above, the SBB keeps its supercritical character at all values of
$q$ for $l=1$. The possibility to switch between the different types of the
SBB, i.e., between the phase transitions of the second and first kinds by
means of the mismatch, may be readily implemented in the experiment, as the
mismatch may be varied by means of the wavelength of the input beam.

The trend of the transition to the subcritical bifurcation with the increase
of $q$ complies with the fact that, as mentioned above, at large $q$ the
cascading approximation applies to Eqs. (\ref{U}), (\ref{V}), making them
asymptotically tantamount to the equation for the single-component (FF)
lattice with the two sites carrying the cubic self-focusing nonlinearity. In
the latter case, the SBB for the localized modes is subcritical for all $%
l\geq 2$ \cite{BB11}.

\begin{figure}[th]
\epsfig{file=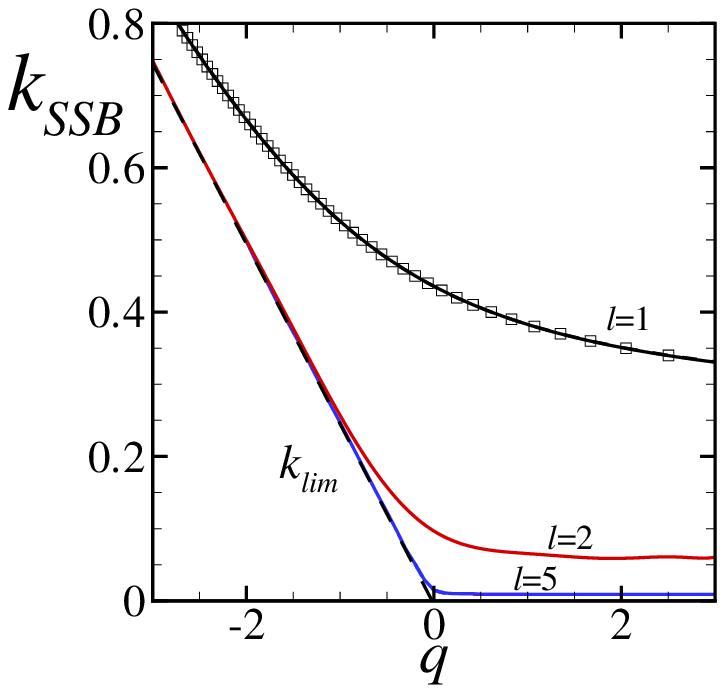,width=6cm}
\caption{(Color online) The critical value of the propagation constant, $k_{%
\mathrm{SBB}}$, at which the symmetry-breaking bifurcation occurs, as a
function of mismatch $q$, at fixed $l=1,2$ and $5$. Open squares for $l=1$
correspond to analytical prediction (\protect\ref{bif-1}). The dashed line
corresponds to $k_{\lim }$, as determined by Eq. (\protect\ref{klim}).}
\label{fig_kSBB_q}
\end{figure}

The results of the analysis performed at all values of mismatch $q$,
negative, zero, and positive, are summarized in Fig. \ref{fig_kSBB_q}, where
the critical value of the propagation constant at which the SBB happens, $k_{%
\mathrm{SBB}}$, is plotted as a function of $q$ at different fixed values of
distance $l$ between the nonlinear sites. Note that, at large $l$, the
bifurcation point $k_{\mathrm{SBB}}$ is very close to the limit value $%
k_{\lim }$ given by Eq. (\ref{klim}), which is simply explained by the fact
that large $l$ implies an exponentially weak coupling between the two
nonlinear sites, hence the stability margin of the symmetric state is
vanishingly small. On the other hand, at large $q>0$ the curves approach
constant values, which correspond, in the cascading limit, to the system
with the cubic nonlinearity.

\section{Conclusion}

The objective of this work is to develop the analysis of the stability and
spontaneous symmetry breaking of discrete solitary modes in one-dimensional
lattices with the quadratic nonlinearity, in the most basic setting with the
nonlinearity applied at a two sites embedded into the linear host lattice,
with distance $l$ between them. The system can be readily implemented as an
optical waveguiding array. The solutions for symmetric, asymmetric and
antisymmetric modes are obtained in the implicit analytical form, which
becomes explicit for $l=1$. The stability of the modes was investigated
through the computation of eigenvalues for small perturbations, and checked
against direct simulations of the evolution of perturbed modes. The analysis
has demonstrated that properties of the modes undergo essential evolution
with the change of the mismatch, $q$: The SBB\ (symmetry-breaking
bifurcation) changes from super- into subcritical (except for the case of $%
l=1$), the existence threshold for the symmetric localized modes vanishes
exactly at the point of $q=0$, and the stability of the branches changes
too. These results suggest a straightforward implementation in the
experiment.

The analysis reported in this work can be naturally extended in other
directions. In particular, following Refs. \cite{BB11} and \cite{2012}, it
may be interesting to consider a finite ring-shaped lattice with the
nonlinear sites set at diametrically opposite points. A challenging problem
is to generalize the analysis for two-dimensional lattices.

\end{document}